\documentclass{PoS}

\usepackage{booktabs}
\usepackage{hepunits}
\usepackage{placeins}

\newcommand{\Nf}{N_{\mathrm f}}

\newcommand{\msbar}{{\overline{\rm MS}}}
\DeclareMathOperator{\Tr}{Tr}

\title{Heavy semileptonics with a fully relativistic mixed action}

\ShortTitle{Heavy semileptonics with a fully relativistic mixed action}

\author{\speaker{J.~Frison}, A. Bussone, G.~Herdo\'iza, C.~Pena, J.\'A.~Romero, J.~Ugarrio\\
        Instituto de F\'isica Te\'orica UAM-CSIC, c/ Nicol\'as Cabrera 13-15, Universidad Aut\'onoma de Madrid, 28049 Madrid, Spain\\

        E-mail: \email{julien.frison@uam.es}}

\abstract{
  The first phase of a heavy quark program based on twisted mass valence quarks has been presented at last years's lattice conference. The CLS $N_f=2+1$ ensembles were used for their fine lattice spacing, while twisting the masses is expected to reduce discretisation errors even further and allow for a fully relativistic calculation. We present our strategy and preliminary results on three point functions, corresponding to $D\to K$ and $D\to\pi$ semileptonic decays. The form factors for $m_u=m_d=m_s$ quark masses obtained as a first step are shown here to be at the percent level
in statistical precision at $q^2=0$.

}
\FullConference{37th International Symposium on Lattice Field Theory - Lattice2019\\
		16-22 June 2019\\
		Wuhan, China}

\begin{document}

\section{Introduction}

Flavour physics has always been an attractive place to look for new physics, in particular because it is the origin of most of the parameters of the Standard Model,
for which one naturally hopes to find a more fundamental explanation. In the last years some promising tensions indeed appeared in $b\to s$ and
$b\to c$ transitions, which might increase in the near future as LHCb and Belle II keep accumulating statistics. In the meantime, the BESIII and CLEO-c experiments are
improving our knowledge of $D$ decays. And even if no new physics ends up being discovered, those studies can at least help to get more and more precise values of the aforementioned parameters. 

As those new experimental results come in, our theoretical computations have to improve too. Lattice field theory has become a very powerful method for this objective, since fine enough lattices to fit
relativistic 
heavy quarks can now be generated. In the last years, this front as well has seen a large increase of the influx of new results \cite{Aoki:2019cca}, but much remains to be done in order to have fully reliable
estimates for all the observables of interest. While most of the previous results are based on some effective heavy quark action,
it has been shown that some quantities can be accessed directly with a fully relativistic action \cite{Carrasco:2013zta,Bussone:2016iua}. This is the approach of the current work. 


In this project we will take advantage of the automatic $O(a)$ improvement \cite{Frezzotti:2003ni} of twisted mass fermions \cite{Frezzotti:2000nk} at maximal twist, which precisely gets rid of the dangerous $O(am_c)$ terms. Unlike the 
strategy of ETMC \cite{Lubicz:2017syv}
however, this action will only be used in the valence, while the $\Nf=2+1$ Wilson action in the sea comes from the very fine lattices obtained by CLS with the help of open boundary conditions \cite{Bruno:2014jqa}. The control
of the order of discretisation errors and the size of the lattice spacing are of course two crucial features when it comes to heavy quark physics. Here those two actions only differ by the choice of mass 
parameters, so the renormalisation factors (in a massless scheme) are kept unchanged and no matching is required except the tuning of the valence masses. This matching has already been described in \cite{Bussone:2019mlt}, as well as first results for leptonic decays in \cite{Bussone:2018wki}.

With this setup, our next step consists in focusing on semileptonic $D\to\pi$ (and $D\to K$) decays to obtain $V_{cd}$ ($V_{cs}$). 
While charm physics studies are relevant by themselves for phenomenology and a better understanding of QCD, they also allow to establish the expected benefits regarding cutoff effects in view
of B-sector computations. 
This paves the way for the later study of processes 
such as $B\to \pi$ ($V_{ub}$), $B\to D^{(*)}$ ($V_{cb}, R(D^{(*)})$), $B\to K^{(*)}$ ($R(K^{(*)}$), as well as potentially $D\to \rho,a_0,f_0$ (for which BESIII published some new experimental results 
\cite{Ablikim:2018qzz}).

\section{Ensembles and correlators}

The CLS ensembles considered in the current scope of this study are described in Tab.~\ref{tab_ens}. 
For the finest of those ensembles, $am_c^{\msbar,2\ \GeV}\sim 0.32$ while $am_b^{\msbar,2\ \GeV}\sim 1.06$. 
As a first step we are going to focus on presenting preliminary results for the $m_u=m_d=m_s$ subset of ensembles, to which we will refer as ``symmetric point''.

\begin{table}[!htbp]
  \begin{center}
    \small
    \begin{tabular}{ccccccccc}
      \toprule
      Id &   $\beta$ & $~a$[fm] & $L/a$  &  $T/a$  & $m_\pi$[MeV] &   $m_K$[MeV] &  $m_\pi L$\\
      \midrule
      \bfseries H101 &\bf 3.40 &\bf 0.087 &\bf 32 &\bf 96 & 420 &420  & 5.8\mdseries\\
      H102 & 3.40 & 0.087 & 32 & 96 & 350 & 440 & 4.9\\
      H105 & 3.40 & 0.087 & 32 & 96     & 280 &460  & 3.9\\
      \midrule
      \bfseries H400 &\bf 3.46 &\bf 0.077 &\bf 32 &\bf 96   & 420 & 420 & 5.2\mdseries\\
      \midrule
      \bfseries H200 &\bf 3.55 &\bf 0.065 &\bf 32 &\bf 96  & 420 & 420 & 4.3\mdseries\\
      N203 & 3.55 & 0.065 & 48 & 128 & 340 & 440 & 5.4\\
      D200 & 3.55 & 0.065 & 64 & 128 & 200 & 480 & 4.2\\
      \midrule
      \bfseries N300 &\bf 3.70 &\bf 0.050 &\bf 48 &\bf 128 & 420 & 420 & 5.1\mdseries\\
      J303 & 3.70 & 0.050 & 64 & 192 & 260 & 260 & 4.1\\
      \bottomrule
    \end{tabular}
    \caption{\label{tab_ens} List of CLS $N_\mathrm{f}=2+1$ ensembles considered in the present study. The second column corresponds to the inverse bare coupling, $\beta=6/g^2_0$. In the third and fourth columns, $L$ and $T$, refer to the spatial and temporal extent of the lattice. Ensembles for which we already have preliminary results to present are in boldface. Approximate values of the pion and Kaon masses are provided~\cite{Bruno:2014jqa,Bruno:2016plf}.}
  \end{center}
\end{table}

On those ensembles we have to compute three-point and two-point functions with momenta. Those are obtained by inversion on a stochastic source $\xi_i$ and the use of a sequential propagator: 
\begin{eqnarray}
Q_{f,i}(x'\mid p) &=& e^{-ipx'}D_f^{-1}(x',x) \xi_i(x)e^{ipx}\delta_{x_0-t_{src}},\quad \textrm{where}\; \sum_i \xi_i(x)\xi_i^*(y)\to \delta(x-y)\\
W_{f'f,i}(x'\mid p',p) &=& e^{-ip'x'}D_{f'}^{-1}(x',x)e^{ip'x}\delta_{x_0-t_{snk}} Q_{f,i}(x\mid p), 
\end{eqnarray}
As a result the correlators are obtained from
\begin{equation}
C^{2pt}_{ff'} = \Tr{\left[ Q_{f} Q_{f'}^\dag \right]}\quad\textrm{and}\quad
C^{3pt,\Gamma}_{ff'f"} = \Tr{\left[ Q_{f} W_{f'f"}^\dag \gamma_5\Gamma\gamma_5 \right]}
\end{equation}
This way we obtain with no extra cost the matrix elements for any $\Gamma$ (in particular the scalar and vector form factors of the Standard Model $D\to\pi$) and can vary as we want the time at 
which this operator is inserted, while the external mesons are restricted to be pseudoscalar for now and are each on a fixed time slice. 

The flavour indices $f,f',f"$ shown here are chosen so that $D\to D$, $D\to P$ ($P=\pi/K$), $P\to D$ and $P\to P$ transitions are all available, which allows to consider various methods to extract the matrix elements and also
allows some sanity checks for this preparatory phase. 
The twisted mass assignment is such that all mesons involved are associated to conserved
currents, thus circumventing issues with $O(a^2)$ flavour breaking effects \cite{Jansen:2005cg,Dimopoulos:2009es}.

We have chosen to keep the momentum of the spectator quark to zero while each of the two other quarks has a momentum which can take 15 values, each one being imposed by a different
twisted boundary condition for the quark inversion \cite{Guadagnoli:2005be}. Those values are positive and negative (7 of each,  going up to $700\ \MeV$) 
because a parity average is necessary to keep the $O(a)$ improvement. As a result we have 15 light inversions and 15 heavy inversions, but this gives us $15^2=225$ kinematics on the correlator, 
covering all the range from $q^2=q_{max}^2$ to a slightly negative $q^2$.

All being considered, this leads to $14430$ correlators per configuration for a single noise hit, a single source-sink separation, degenerate light quarks (u,d,s) and a single choice of 
charm mass (tuned to its physical mass, while several values in $[m_c,m_b]$ will be considered in the future). Despite
this impressive number, we have checked that the cost of contractions remains negligible.

\section{Parametrisation and useful combinations}

The scalar and vector form factors, following arguments of Lorentz symmetry, have the structure: 
\begin{eqnarray}
\langle S\rangle &=& \frac{M_D^2-M_P^2}{\mu_c-\mu_q} f_0(q^2) + {\cal O}(a^2)\\
\langle \hat V_\mu \rangle &=& P_\mu f_+(q^2) + q_\mu \frac{M_D^2-M_P^2}{q^2} \left[ f_0(q^2)-f_+(q^2) \right] + {\cal O}(a^2)
\end{eqnarray}
where $P_\mu=p_{D\mu}+p_{P\mu}$, $q_\mu=p_{D\mu}-p_{P\mu}$, $S$ is unrenormalised and $\hat V=Z_V V$ is renormalised. 

To obtain the matrix elements needed for the form factors, we use the double ratios introduced in \cite{Carrasco:2016kpy}:
\begin{equation}
\mid\langle\hat V_\mu\rangle\mid^2 = 4p_{D\mu}p_{P\mu} \frac{C^{3pt,\gamma_\mu}_{DP}(t,t')C^{3pt,\gamma_\mu}_{PD}(t,t')}{C^{3pt,\gamma_\mu}_{DD}(t,t')C^{3pt,\gamma_\mu}_{PP}(t,t')}
\qquad
\mid\langle S\rangle\mid^2 = 4E_DE_P \frac{C^{3pt,\gamma_5}_{DP}(t,t')C^{3pt,\gamma_5}_{PD}(t,t')}{C^{2pt}_D(t')C^{2pt}_P(t')} .
\end{equation}
Those double ratios indeed do not depend on the result of a fit, can benefit from statistical cancellations of correlated terms, do not need an explicit renormalisation, and, 
more importantly, look more robust against excited states contaminations than the naive ratios of three-point to two-point functions. We can notice that, in the approximation
$\langle M_1,n_1\mid\Gamma\mid M_2, n_2\rangle \approx \langle M_1,0\mid\Gamma\mid M_2,0\rangle$ in which the form factors do not change much when involving higher and higher excited states,
the cancellation of excited states in the double ratio method is exact. 

We also describe the hypercubic discretisation effects in terms of the invariants used by \cite{Lubicz:2017syv}, which presented them as a major worry:
\begin{eqnarray}
\langle S\rangle_{\rm hyp} &=& \frac{a^2}{\mu_c-\mu_q}\left[q^{[4]}\tilde H_1 + q^{[3]}P^{[1]}\tilde H_2 + q^{[2]}P^{[2]}\tilde H_3 + q^{[1]}P^{[3]}\tilde H_4+ P^{[4]}\tilde H_5 \right] \\
\langle \hat V_\mu\rangle_{\rm hyp} &=& a^2 \left[ (q_\mu)^3 H_1 + (q_\mu)^2P_\mu H_2 + q_\mu (P_\mu)^2 H_3 + (P_\mu)^3 H_4 \right] .
\end{eqnarray}
but we will actually exploit this structure more intensively by using non-democratic momenta. Democratic (four-)momenta are often chosen to
minimise the ${\cal O}(q^{[4]})$ discretisation effects, however here the symmetry is already strongly broken by a fundamentally different treatment of time and space, which will likely be the dominant source
of imbalance between components. For simplicity we choose to nevertheless keep the momentum on a fixed line, in which case $q_i/P_i$ does not depend on the spatial $i$ and
we can easily form interesting model-independent combinations based on
\begin{eqnarray}
\langle \hat V_i\rangle_{\mathrm{hyp}, jk} &\equiv& \left[\langle \hat V_j\rangle/q_j - \langle \hat V_k\rangle/q_k\right]q_i^3/(q_j^2-q_k^2)
\end{eqnarray}
or if we want to perform an analysis ignoring hypercubic effects we can directly build the combination adding in quadrature (instead of naively like for democratic momenta)
\begin{equation}
\langle \hat V_l\rangle_{\rm avg} \equiv \left[ \sum_{i} \frac{q_l}{q_i} \frac{ \langle \hat V_i\rangle}{\sigma(V_i)^2} \right]
           / \left[ \sum_{i} \frac{1}{\sigma(V_i)^2} \right]
\end{equation}

Finally, let us remind that those hypercubic effects are also probed by the breaking of the Ward-Takahashi Identity (WTI) as
\begin{equation}
\Delta_{\rm WTI}^{\rm hyp} = (\mu_c-\mu_q)\langle S\rangle + q_\mu\langle\hat V_\mu\rangle = (\mu_c-\mu_q)\langle S\rangle_{\rm hyp} + q_\mu\langle\hat V_\mu\rangle_{\rm  hyp} .
\end{equation}

All those relations do not provide enough constraints to solve for all the continuum and hypercubic form factors at every kinematical point, but they give us a very strong control on
the estimate of those sources of error, and do give us direct access to the Lorentz-invariant $f_+(q^2)$ (resp. $f_-(q^2)$) in the $p_D=p_P$ (resp. $p_D=-p_P$) frame.

\section{Results}

\begin{figure}[ht]
\begin{center}
    \includegraphics[width=0.225 \textwidth]{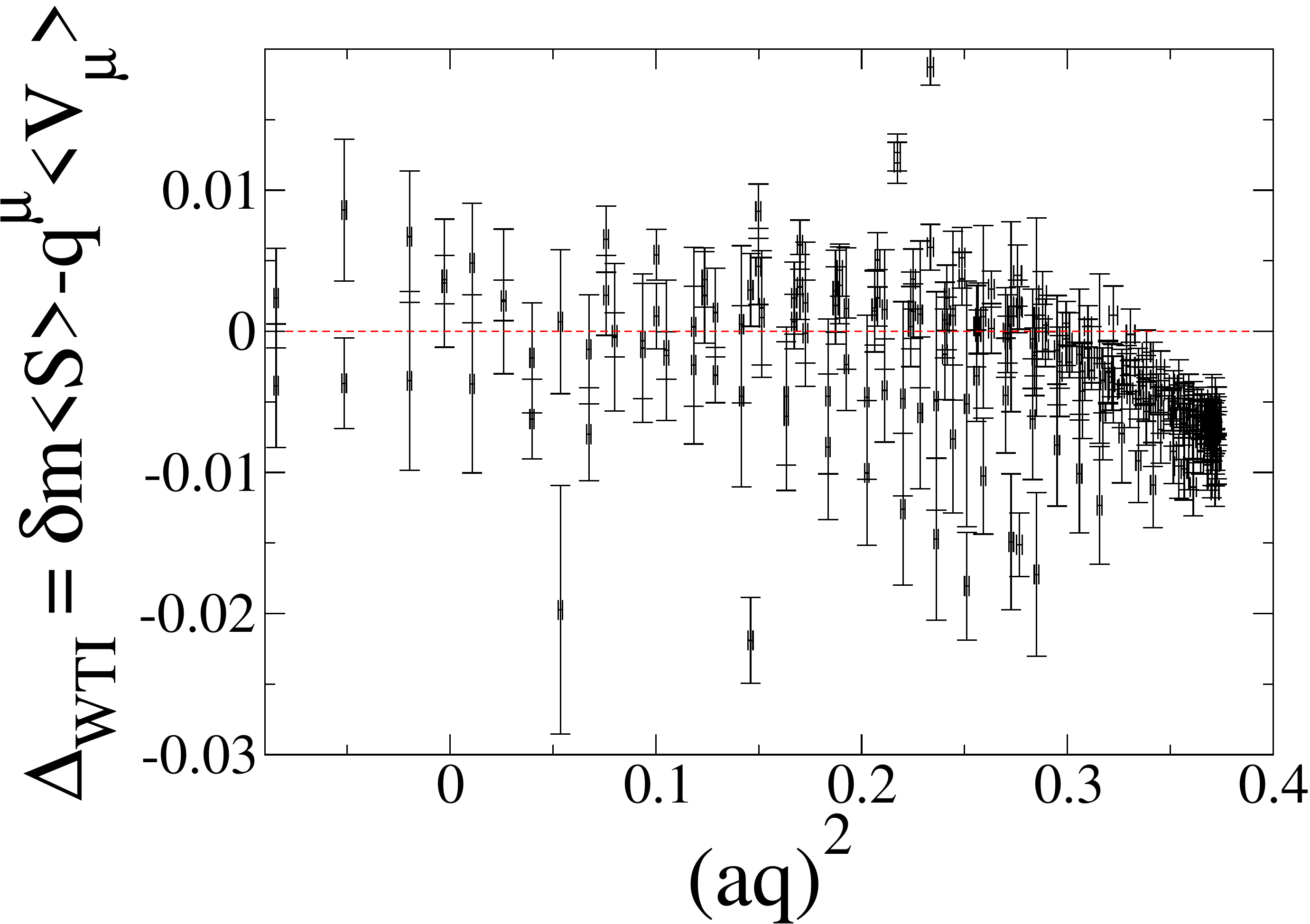}
    \hspace{0.5cm}
    \includegraphics[width=0.2 \textwidth]{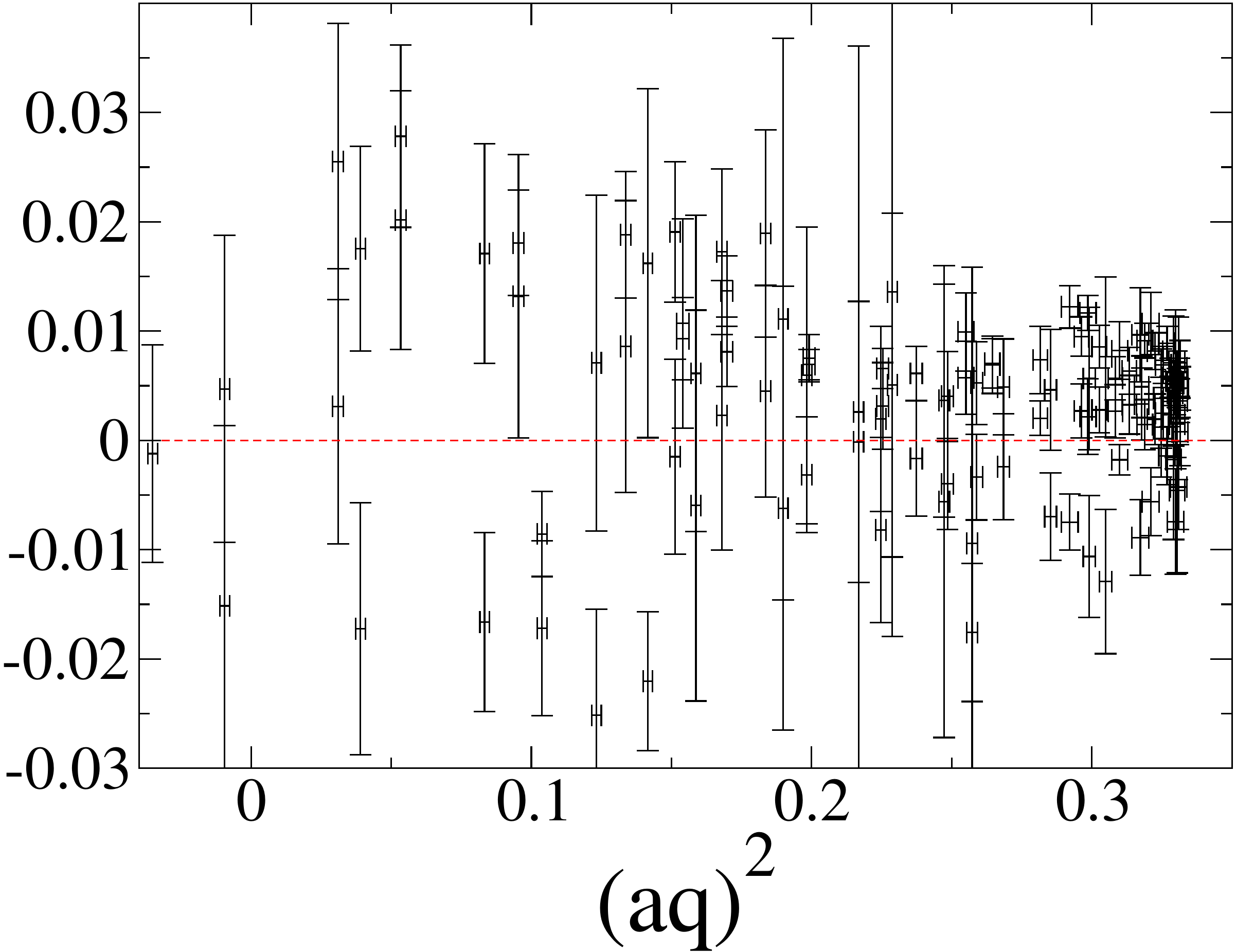}
    \hspace{0.5cm}
    \includegraphics[width=0.2 \textwidth]{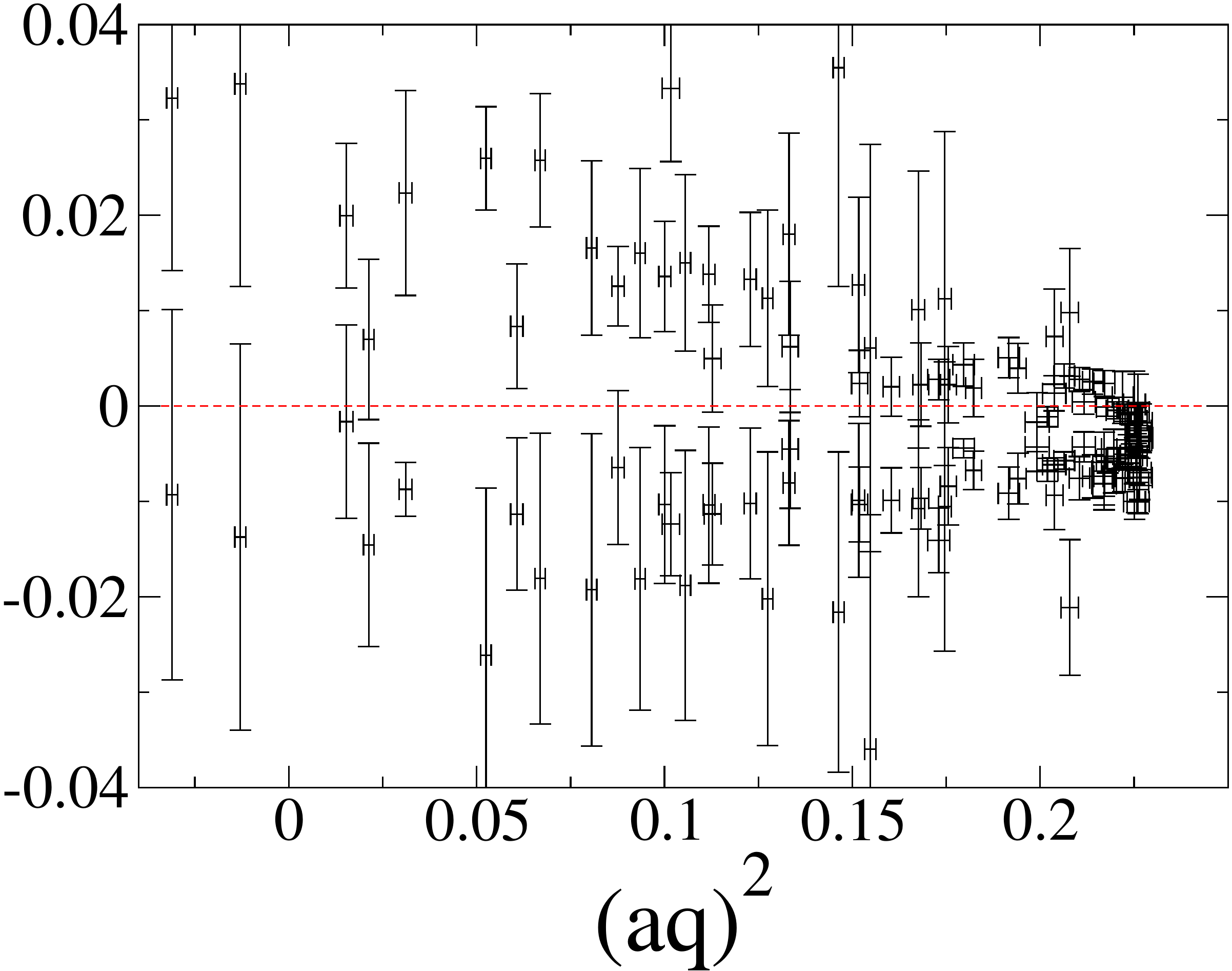}
    \hspace{0.5cm}
    \includegraphics[width=0.2 \textwidth]{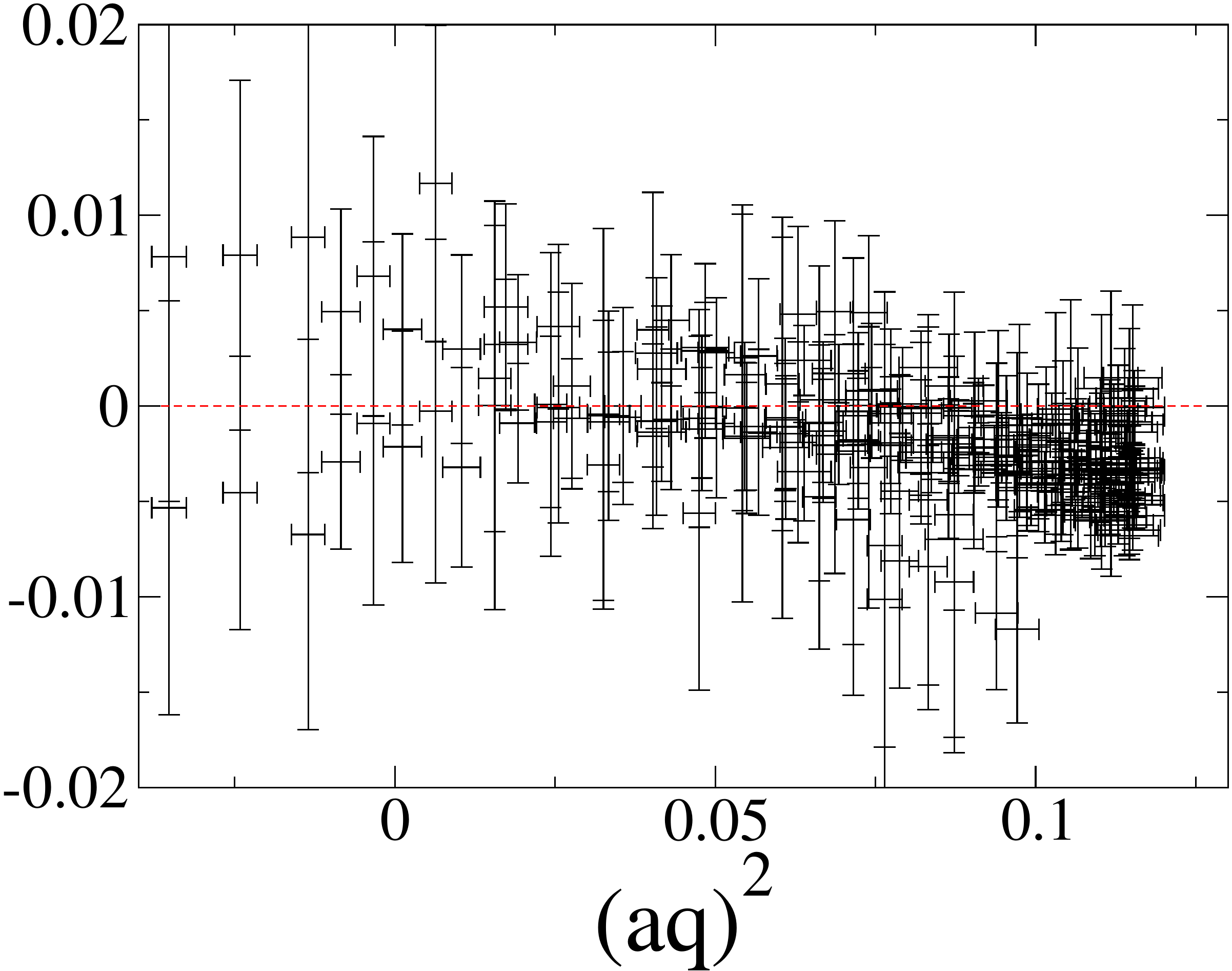} 
  \caption{Violation of the Ward identity $\Delta^{\rm hyp}_{\rm WTI}$ shown with H101 (coarsest, left), H400, H200, N300 (finest, right). 
It is small, with little signal, and -apart from a slight tendency in the coarsest case- they do not form any coherent trend as a function of $q^2$ (or any other hypercubic invariant). }
  \label{fig:WTI}
\end{center}
\end{figure}

\begin{figure}[ht]
\begin{center}
  \begin{tabular}{c}
    \includegraphics[width=0.225 \textwidth]{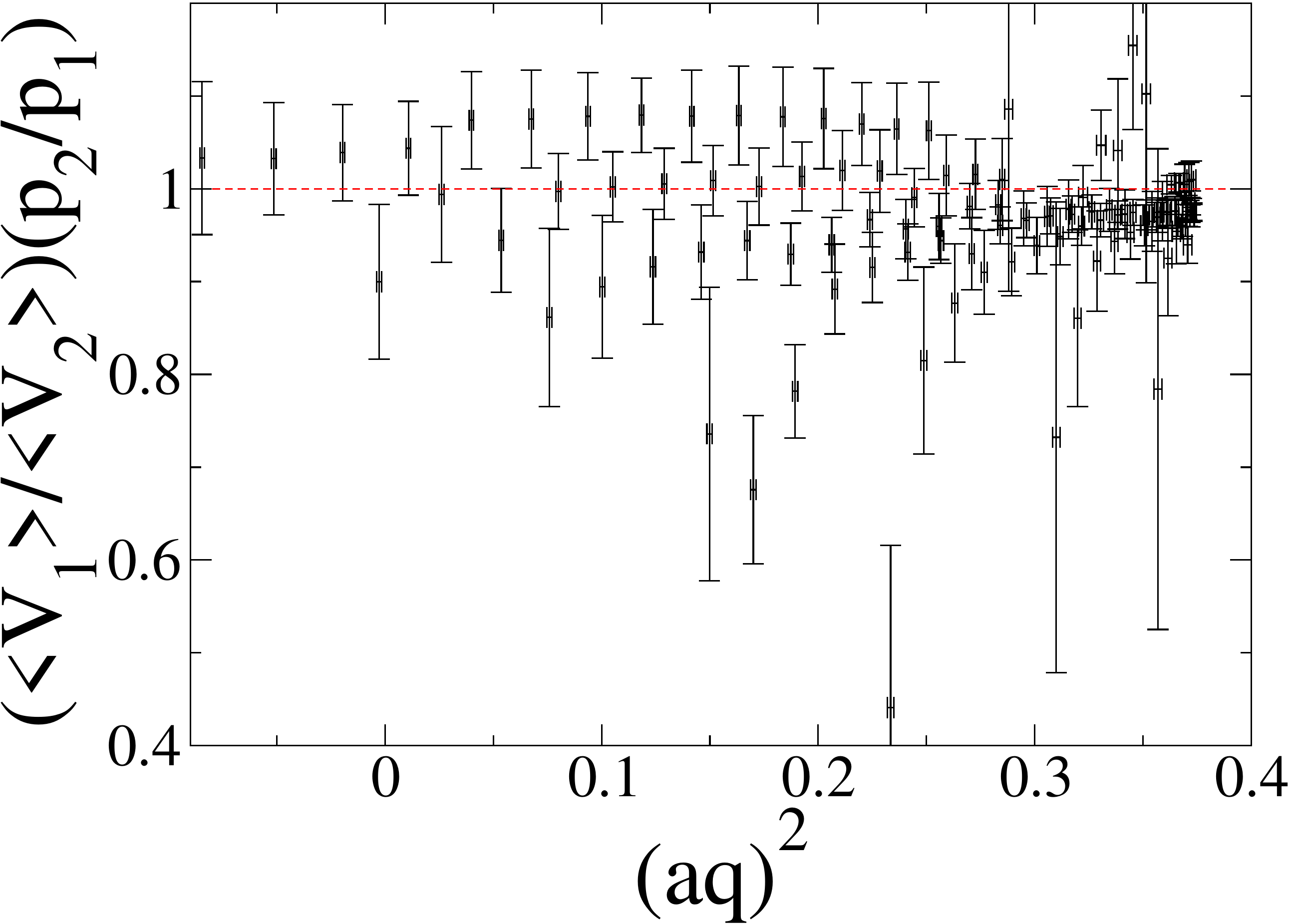}
    \hspace{0.5cm}
    \includegraphics[width=0.2 \textwidth]{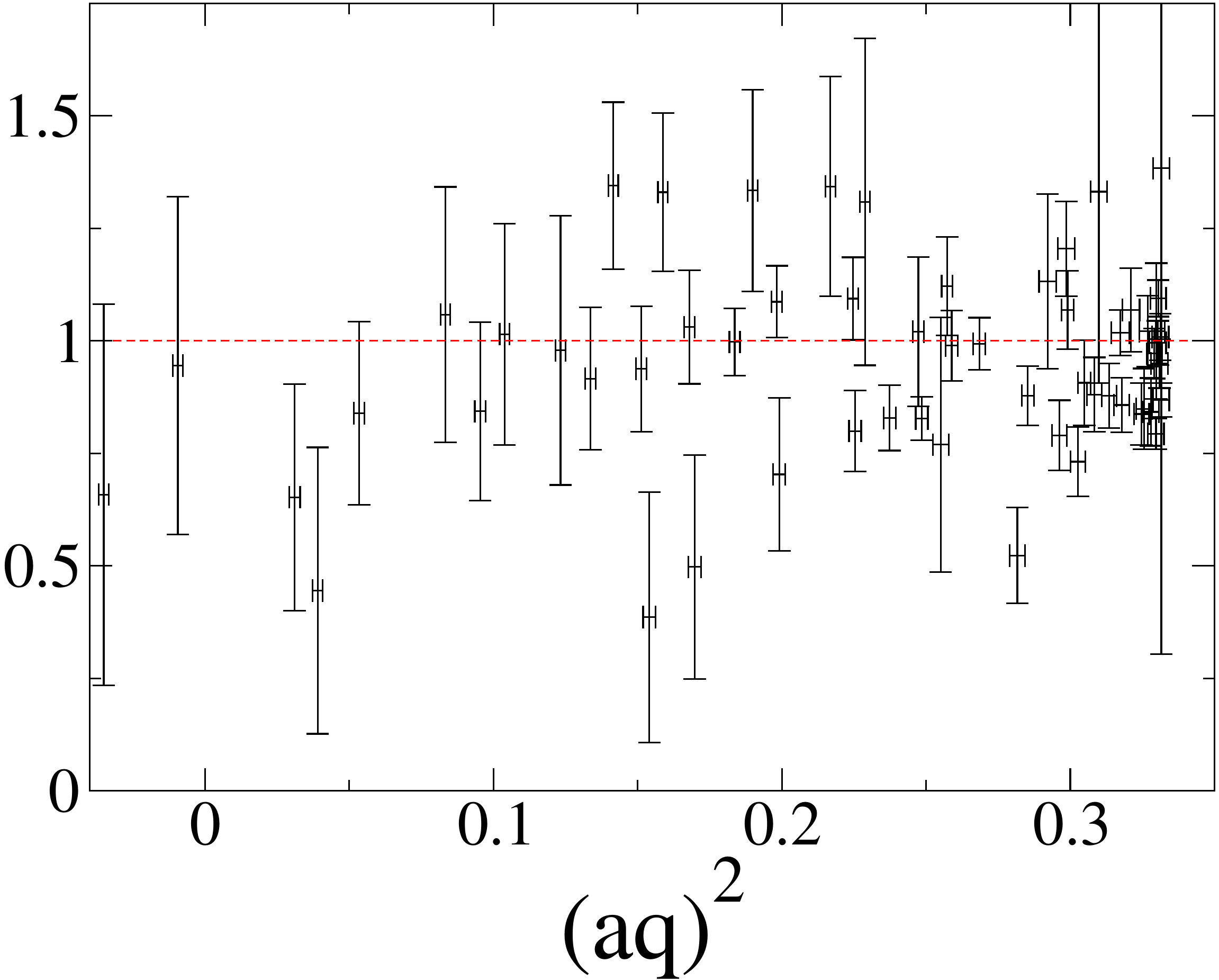}
    \hspace{0.5cm}
    \includegraphics[width=0.2 \textwidth]{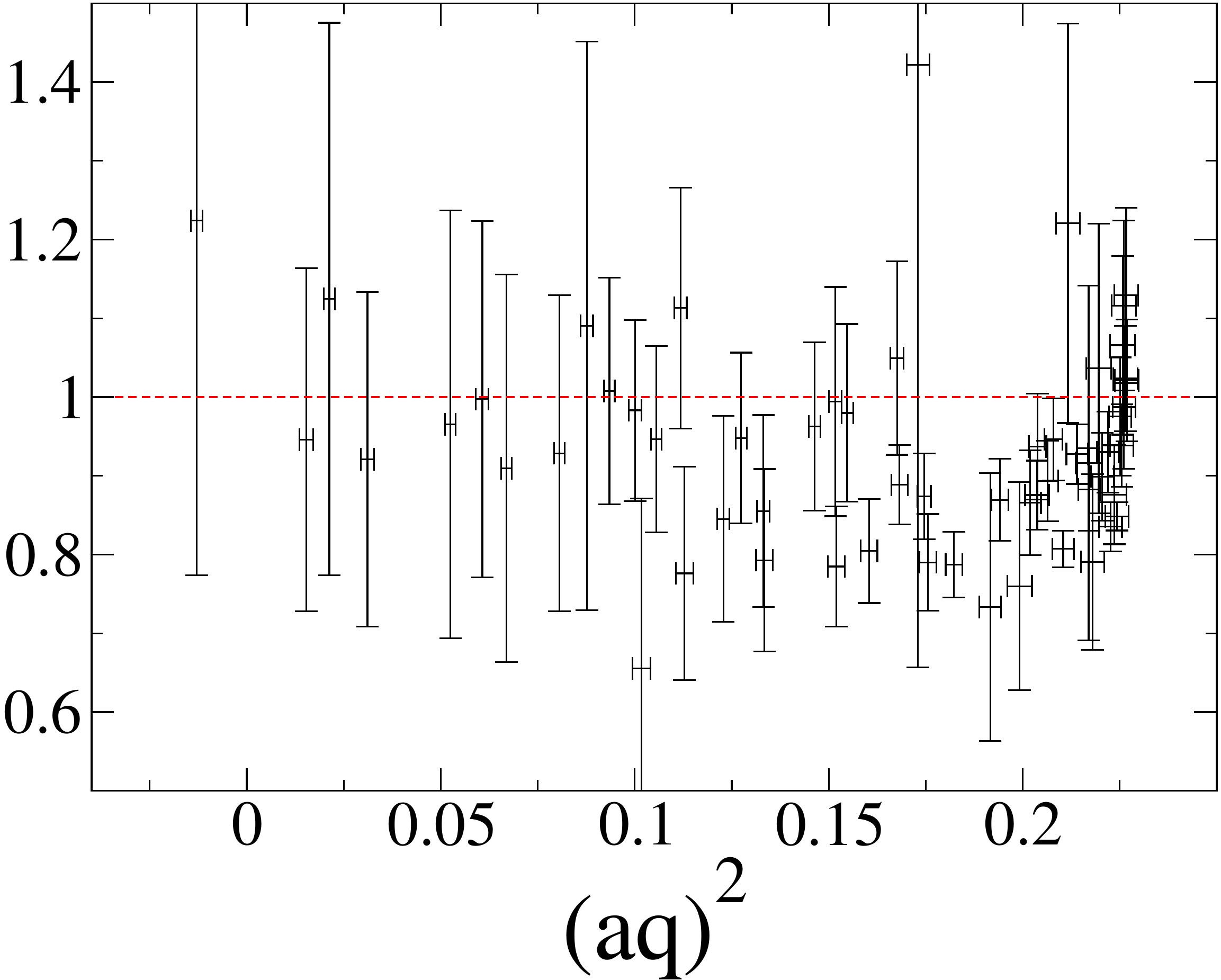}
    \hspace{0.5cm}
    \includegraphics[width=0.2 \textwidth]{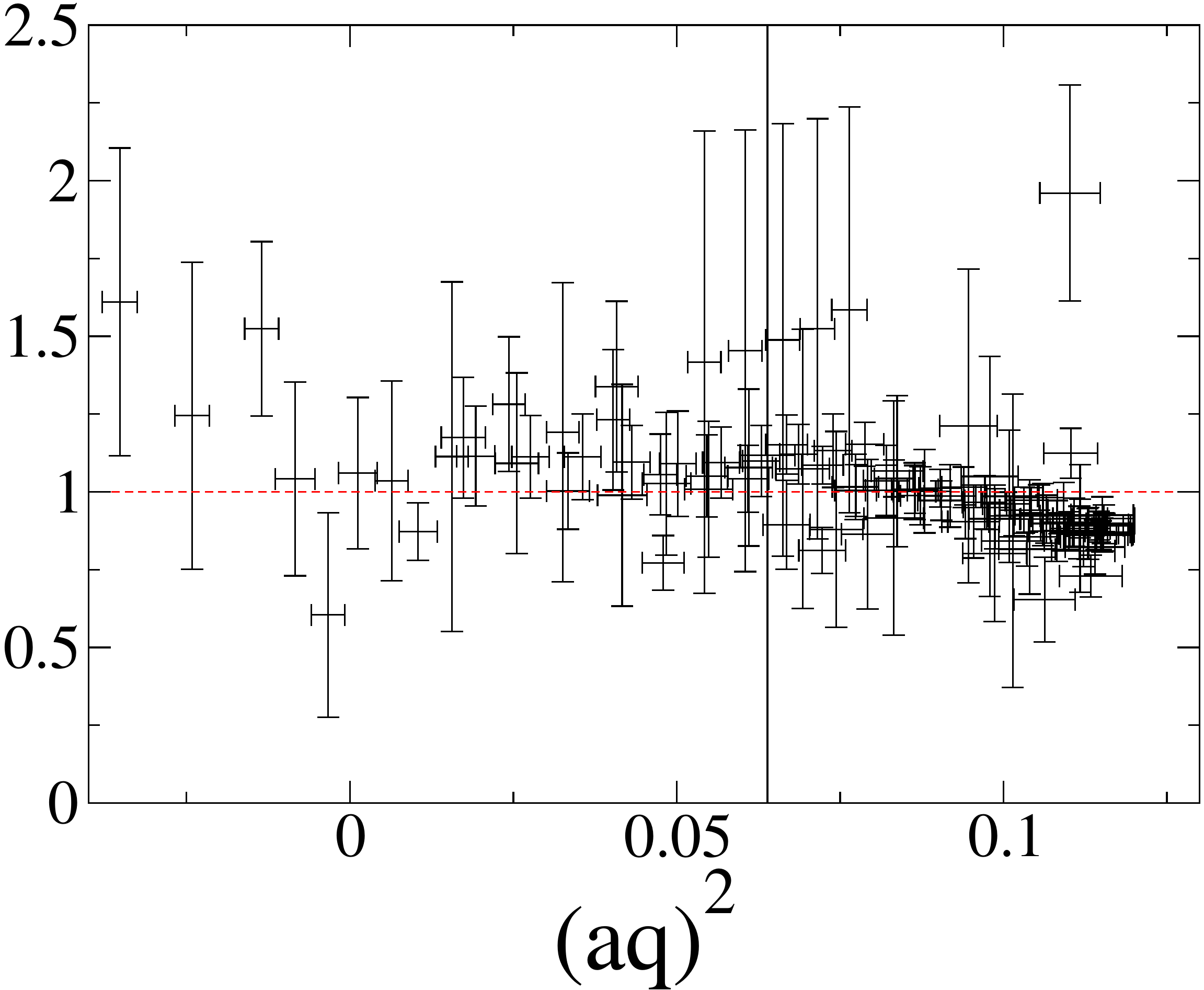} 
  \end{tabular}
  \caption{These ratios of spatial vector matrix elements are more sensitive to hypercubic effects than the Ward identity (dominated by the time component), in particular for the coarsest ensemble. }
  \label{fig:RVratios}
\end{center}
\end{figure}

We now present a preliminary analysis at the symmetric point. Let us first look at the WTI (Fig.~\ref{fig:WTI}) and the spatial hypercubic effects (Fig.~\ref{fig:RVratios}): we notice 
that hypercubic effects on those observables
appear to be relatively small. The values seem scattered randomly and no clear dependence emerges by choosing any other hypercubic invariant as the x axis, contrary to the dominance of an ${\cal O}(q^{[4]})$ term which
was observed in \cite{Lubicz:2017syv}. This justifies neglecting the hypercubic effects in the current preliminary analysis, while showing the potential of non-democratic momenta to build sensitive
quantities from which hypercubic form factors could be fitted in a later stage of our study. 

\begin{figure}[ht]
\begin{center}
  \includegraphics[width=0.6 \textwidth]{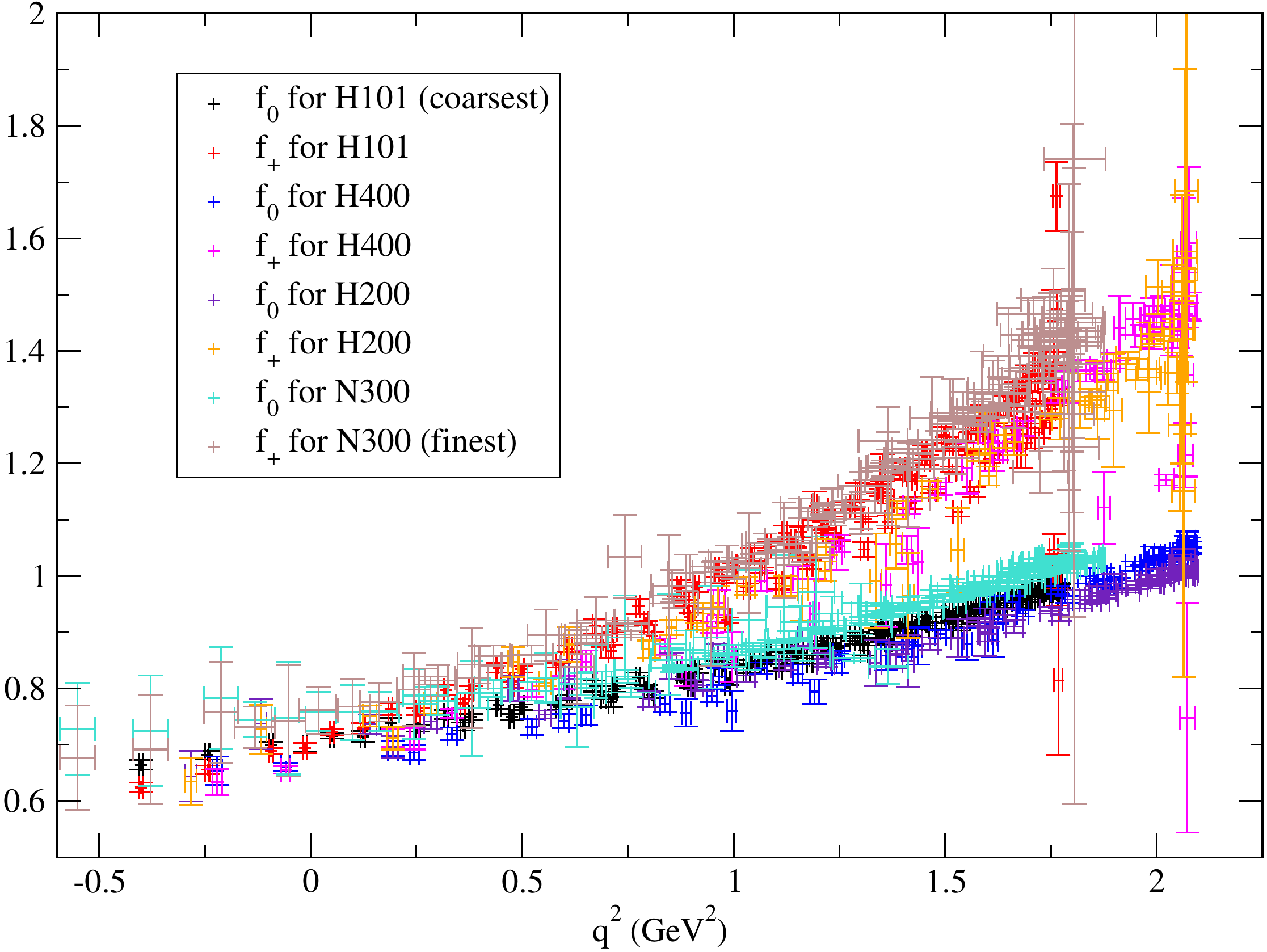}
  \caption{While some hypercubic effects are visible for coarse ensembles at large $q^2$ (where many different kinematics are available), most points are on a smooth line and the results for
H101, H200, H400 are close. The small differences are certainly a mix of discretisation effects and mass matching, which can not be disentangled at this stage of the analysis. 
Hypercubic effects are neither visible around $q^2=0$, nor on $f_0(q^2)$, which is mostly determined by $\langle S\rangle$. $f_+(q^2)$ is more sensitive because it depends more strongly on
$\langle V_\mu\rangle$, but it is also the case in which we can expect the greatest gain from the use of non-democratic momenta once a global fit is performed. }
  \label{fig:f0fp}
\end{center}
\end{figure}

We then solve for the Lorentz-invariant form factors $f_0(q^2),f_+(q^2)$ (still containing ${\cal O}(a^2,[ap]^2)$ effects) from the data $(\langle S\rangle,\langle V_\mu\rangle)$. 
Uncorrelated $\chi^2$ are very small, which corresponds to the redundancy of information when hypercubic effects are negligible. 
The results in Fig.~\ref{fig:f0fp} show smooth curves with very little of the ``sawtooth'' behaviour characteristic of hypercubic effects, and we can observe a high level of compatibility
between the results at different lattice spacings. 

As an indication of the potential of our approach, we provide a preliminary determination for the continuum limit of the form factor at the symmetric point, computed directly at $q^2=0$, with an incomplete account of the systematics:
\begin{equation}
f_+(0) = 0.700 (8)_{\rm stat}(23)_{\rm cont}.
\end{equation}
This value is of little interest per se, but gives an idea of the level of precision we can hope for regarding the forthcoming analysis of ensembles with near-physical light quark mass.
Indeed this looks promising compared to the error bar of the current state-of-the-art physical result \cite{Aoki:2019cca}
\begin{equation}
f_+^{D\pi}(0) = 0.666(29)\quad\textrm{and}\quad
f_+^{DK}(0) = 0.747(19)
\end{equation}

\section{Conclusion}

We have shown that our mixed action performs well on charm semi-leptonics, so that dealing with discretisation effects should be only a minor hassle. We have also introduced
non-democratic momenta as a new tool to monitor those discretisation effects, and discussed how it could be used to remove hypercubic breaking effects. Our strategy has been
applied to ensembles with $m_u=m_d=m_s$ of various lattice spacings, giving a very satisfying level of precision on the unphysical symmetric form factors at heavy pion mass. 

The next step of this study will be adding ensembles with various pion masses and lattice volumes. Some data has already been accumulated for that. It will provide
a better control on two sources of systematic errors: masses mistuned slightly away from the line of constant physics, and finite volume errors (with their own hypercubic effects). More importantly, 
analysing the data with near-physical pion mass will also allow to extrapolate the form factors to phenomenologically relevant quantities, 
which will impact the determination of the CKM elements $|V_{cd}|$ and $|V_{cs}|$.

An improvement of the statistics might be considered for some of the noisiest ensembles, in order to have a good control of the systematics. We still have plenty of room for that,
without having to generate more configurations, since only one stochastic noise has been used in the present analysis and we have good reasons to expect that we are far from saturation. 
We also expect an improvement of precision once our analysis fully exploits the momentum dependence.

In the future, the good behaviour of this action for charm semi-leptonics paves the way for an increase of the heavy quark mass towards a study of B decays. This could also benefit from the even
finer ensembles which are being generated by CLS. 

\section{Acknowledgments}
We thank the CLS effort for the generation of the gauge configuration ensembles
used in this study. We acknowledge PRACE (project HFlavLat) and RES access to MareNostrum
at the Barcelona Supercomputing Center (BSC), Spain. We thank CESGA for granting
access to FinisTerrae II. 
We thankfully acknowledge support through the Spanish projects FPA2015-68541-P
(MINECO/FEDER) and PGC2018- 094857-B-I00, the Centro de Excelencia Severo Ochoa
Programme SEV-2016-0597, the EU H2020-MSCA- ITN-2018-813942 (EuroPLEx), and the Ram\'on y Cajal Programme RYC-2012-10819.



\end{document}